\documentclass[a4paper,12pt]{article}
\usepackage[T1]{fontenc}
\usepackage{amsfonts}
\usepackage{ae,aecompl}
\usepackage{amsmath}
\usepackage{amssymb}
\usepackage{amsthm}
\usepackage[pdftex]{graphicx}
\usepackage{pstricks}
\usepackage{tikz}
 \usepackage{epstopdf}
\usepackage[small]{caption}   
\usepackage{color}
\usepackage{enumerate} 
\usepackage{floatrow}
\usepackage{hyperref}
\usepackage{mathdots}
\usepackage{hyperref}
\hypersetup{
    colorlinks=true,
    linkcolor=blue,
    filecolor=magenta,      
    urlcolor=blue,
}

\newcommand{\ie}[0]{\textit{i.e.}}

\definecolor{lightgray}{gray}{0.8}

\usepackage{media9}

\title{Tomographic X-ray data of 3D emoji}
\author{A. Meaney\footnote{Department of Mathematics and Statistics, University of Helsinki, Finland (alexander.meaney@helsinki.fi)},
Z. Purisha\footnote{Department of Mathematics and Statistics, University of Helsinki, Finland (zenith.purisha@helsinki.fi)}  
\ and S. Siltanen\footnote{Department of Mathematics and Statistics, University of Helsinki, Finland (samuli.siltanen@helsinki.fi)}}

\begin{document}

\maketitle

\abstract{This is the documentation of the tomographic X-ray data of emoji phantom made available \href{https://zenodo.org/record/1183532#.WpA35Y5rIy1}{here}. The data can be freely used for scientific purposes with appropriate references to the data and to this document in \url{http://arxiv.org/}. The data set consists of (1) the X-ray sinogram of a single 2D slice of 33 emoji faces (contains 15 different emoji faces) made by small squared ceramic stones and (2) the corresponding static and dynamic measurement matrices modeling the linear operation of the X-ray transform. Each of these sinograms was obtained from a measured 60-projection fan-beam sinogram by down-sampling and taking logarithms. The original (measured) sinogram is also provided in its original form and resolution. The original (measured) sinogram is also provided in its original form and resolution.}

\section{Introduction}

\begin{figure}
\begin{picture}(390,500)
\put(0,500){\includegraphics[width=100pt]{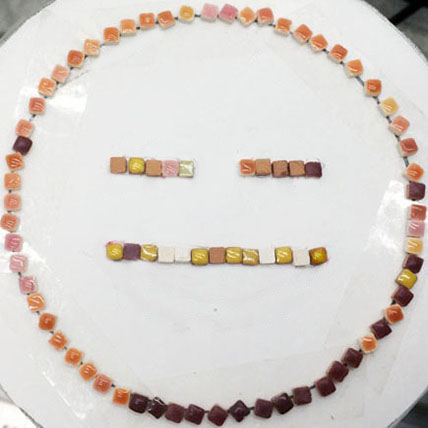}}
\put(150,500){\includegraphics[width=100pt]{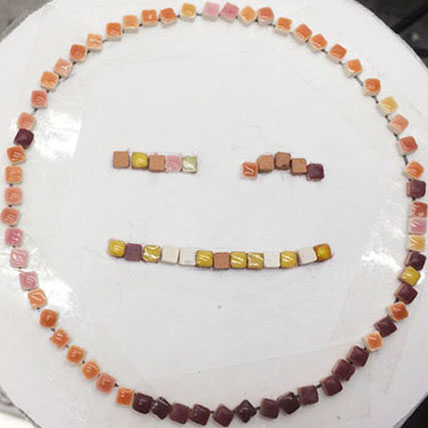}}
\put(300,500){\includegraphics[width=100pt]{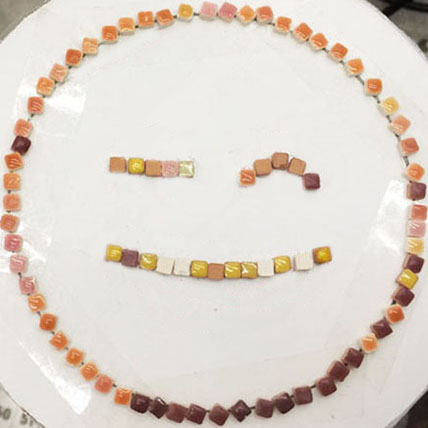}}
\put(0,380){\includegraphics[width=100pt]{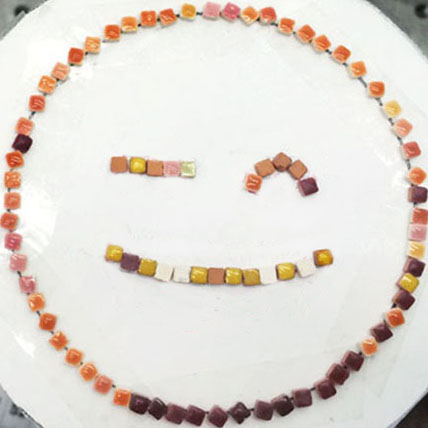}}
\put(150,380){\includegraphics[width=100pt]{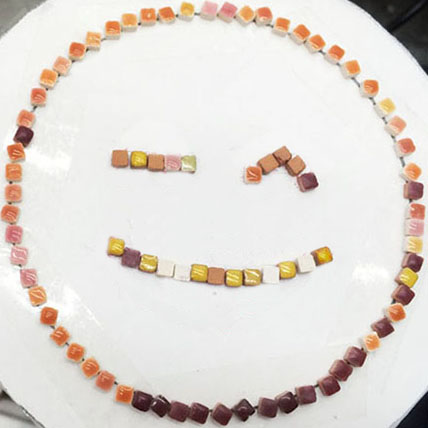}}
\put(300,380){\includegraphics[width=100pt]{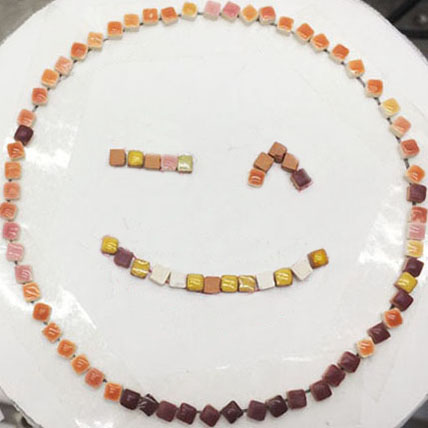}}
\put(0,260){\includegraphics[width=100pt]{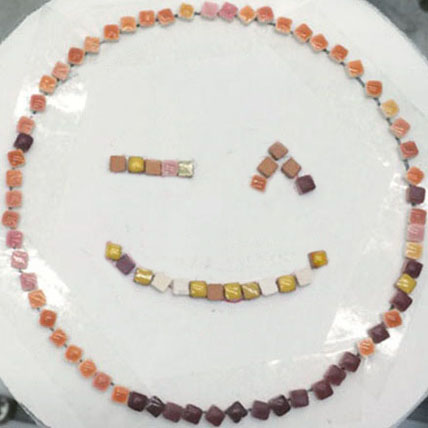}}
\put(150,260){\includegraphics[width=100pt]{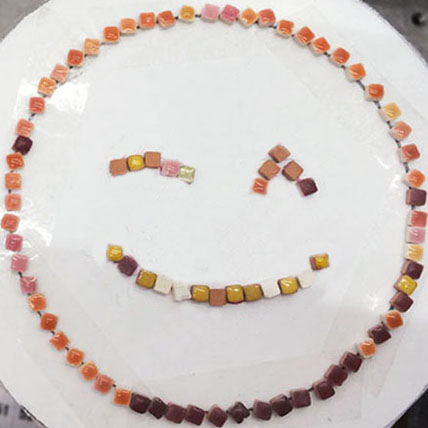}}
\put(300,260){\includegraphics[width=100pt]{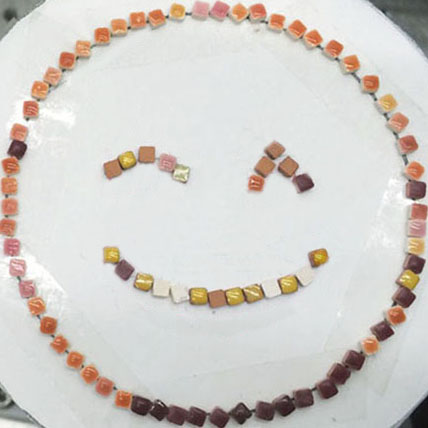}}
\put(0,140){\includegraphics[width=100pt]{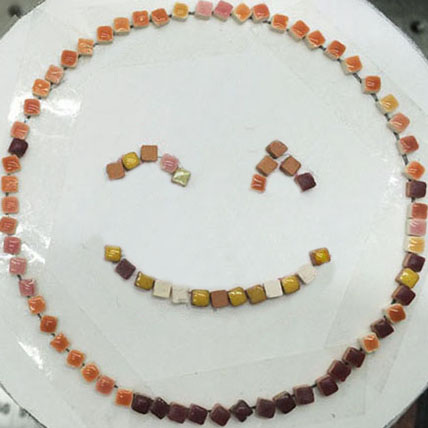}}
\put(150,140){\includegraphics[width=100pt]{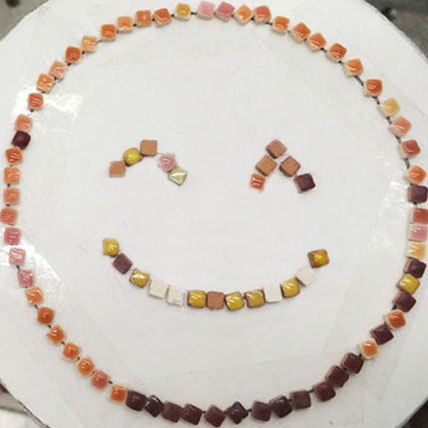}}
\put(300,140){\includegraphics[width=100pt]{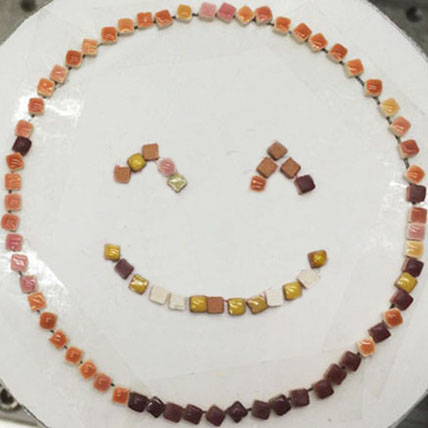}}
\put(0,20){\includegraphics[width=100pt]{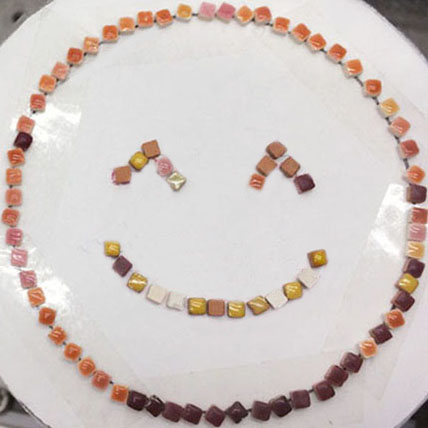}}
\put(150,20){\includegraphics[width=100pt]{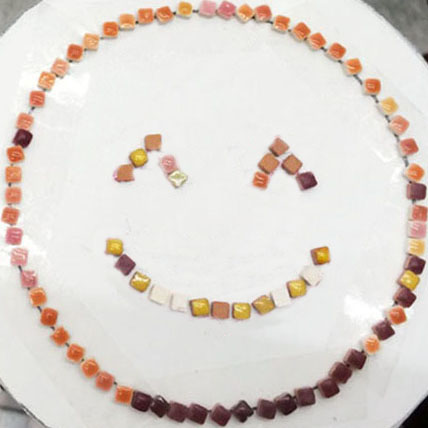}}
\put(300,20){\includegraphics[width=100pt]{frame15c}}
\put(42,490){(1)}
\put(192,490){(2)}
\put(342,490){(3)}
\put(42,370){(4)}
\put(192,370){(5)}
\put(342,370){(6)}
\put(42,250){(7)}
\put(192,250){(8)}
\put(342,250){(9)}
\put(42,130){(10)}
\put(192,130){(11)}
\put(342,130){(12)}
\put(42,10){(13)}
\put(192,10){(14)}
\put(342,10){(15)}
\end{picture}
\caption{Photographs of 15 emoji faces built from ceramic stones. }\label{fig:TargetEmoji}
\end{figure}

The main idea behind the project was to create real CT measurement data for testing sparse-data tomography algorithms. Fifteen emoji faces (frames) which contain circular boundary, mouth and eyes made by small squared ceramic stones of about $3$ mm $\times$ $3$ mm have been designed to recover the evolving emoji faces in every time step. The target is evolving in time from the expressionless face, namely a face with narrow, closed eyes and a straight mouth (photograph (1) in Figure~\ref{fig:TargetEmoji}), to a smiling face with smiling eyes (photograph (15) in Figure~\ref{fig:TargetEmoji}). In the intermediary frames, as it can be seen in Figure~\ref{fig:TargetEmoji}), the mouth evolves from the straight line to a convex parabola, and the eyes, one by one turns from a straight line to a caret/circumflex shape.  The non-stationary target is challenging for sparse dynamic tomography applications. 
The CT data in this data set has been used for testing a shearlet-based sparsity-promoting reconstruction method, see \cite{bubba2017shearlet}. Demonstration of dynamic emoji phantom tomography with several methods is available \href{https://www.youtube.com/watch?v=JTdVAQTFKxI}{here}.

\section{Contents of the data set}\label{sec:datasets}

The data set contains the following MATLAB\footnote{MATLAB is a registered trademark of The MathWorks, Inc.} data files:
\begin{itemize}
\item  {\tt DataStatic\_128x60.mat},
\item  {\tt DataStatic\_128x30.mat},
\item  {\tt DataDynamic\_128x60.mat},
\item  {\tt DataDynamic\_128x30.mat},
\end{itemize}
All the datasets contain CT sinograms and the corresponding measurement matrices with the same resolutions $128 \times 128$ as spatial resolution and $33$ as a temporal resolution in 3D. The 33 times instances are obtained by doubling each of the 15 emoji faces and the last emoji face is added three more times. The data in files {\tt DataStatic\_128x60.mat} and {\tt DataStatic\_128x30.mat} lead to reconstructions with the same geometry in every time instance. In files {\tt DataDynamic\_128x60.mat} and {\tt DataDynamic\_128x30.mat}, the projection angles shift by one degree in every time step, for example, if the angles for time $t_1$ are $[1\quad13\quad25 \,\cdots]$, then for time $t_2$, the angles are $[2\quad14\quad26 \,\cdots]$. Detailed contents of each data file below.

\bigskip\noindent
{\tt DataStatic\_128x60.mat} contains the following variables:
\begin{enumerate}
\item Sparse matrix {\tt A} of size $429\,660\times 540\,672$; measurement matrix.
\item Matrix {\tt sinogram} of size $217\times 1980$; sinogram (60 projections out of full 360 degree circle).
\end{enumerate}

\bigskip\noindent
{\tt DataStatic\_128x30.mat} contains the following variables:
\begin{enumerate}
\item Sparse matrix {\tt A} of size $214\,830\times 540\,672$; measurement matrix.
\item Matrix {\tt sinogram} of size $217\times 990$; sinogram (30 projections out of full 360 degree circle).
\end{enumerate}

\bigskip\noindent
{\tt DataDynamic\_128x60.mat} contains the following variables:
\begin{enumerate}
\item Sparse matrix {\tt A} of size $429\,660\times 540\,672$; measurement matrix.
\item Matrix {\tt m} of size $217\times 1980$; sinogram (60 projections).
\end{enumerate}

\bigskip\noindent
{\tt DataDynamic\_128x30.mat} contains the following variables:
\begin{enumerate}
\item Sparse matrix {\tt A} of size $214\,830\times 540\,672$; measurement matrix.
\item Matrix {\tt sinogram} of size $217\times 990$; sinogram (30 projections).
\end{enumerate}

\bigskip\noindent
Details on the X-ray measurements are described in Section \ref{sec:Measurements} below.
The model for the CT problem is
\begin{equation}\label{eqn:Axm}
 {\tt A*x=m(:)},
\end{equation}
where {\tt m(:)} denotes the standard vector form of matrix {\tt m} in MATLAB ({\tt m} corresponds to matrix {\tt sinogram}) and {\tt x} is the reconstruction in vector form. In other words, the reconstruction task is to find a vector {\tt x} that (approximately) satisfies \eqref{eqn:Axm} and possibly also meets some additional regularization requirements.
A demonstration of the use of the data is presented in Section \ref{sec:Demo} below.

\section{X-ray measurements}\label{sec:Measurements}
\begin{figure}
\begin{picture}(390,300)
\put(0,0){\includegraphics[width=400pt]{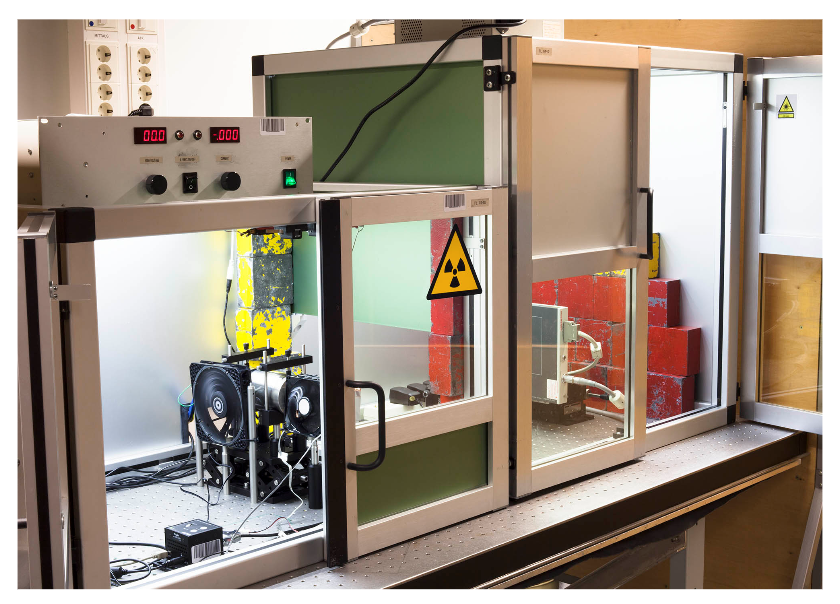}}
\end{picture}
\caption{The custom-made measurement device at University of Helsinki.}\label{fig:CTmachine}
\end{figure}

The data in the sinograms are X-ray tomographic (CT) data of a 2D cross-section of the emoji built from ceramic stones measured with a custom built $\mu$CT device shown in Figure~\ref{fig:CTmachine}. 
\begin{itemize}
\item The X-ray tube is a model XTF5011 manufactured by Oxford Instruments. This model is no longer sold by Oxford Instruments, although they have newer, similar models available. The tube uses a molybdenum ($Z = 42$) target. 
\item The rotation stage is a Thorlabs model CR1/M-27.
\item The flat panel detector is a Hamamatsu Photonics C7942CA-22. The active area of the 
at panel detector is $120$ mm $\times 120$ mm. It consists of a $2400 \times 2400$ array of $50$ $\mu$m pixels. According to the manufacturer, the number of active pixels is  $2240 \times 2344$. However, the image files actually generated by the camera were $2240 \times 2368$ pixels in size.
\end{itemize}

\begin{figure}
\begin{picture}(390,250)
\put(0,0){\includegraphics[width=220pt]{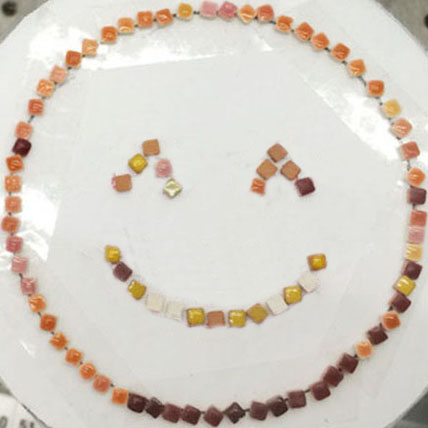}}
\put(240,-14){\includegraphics[height=130pt]{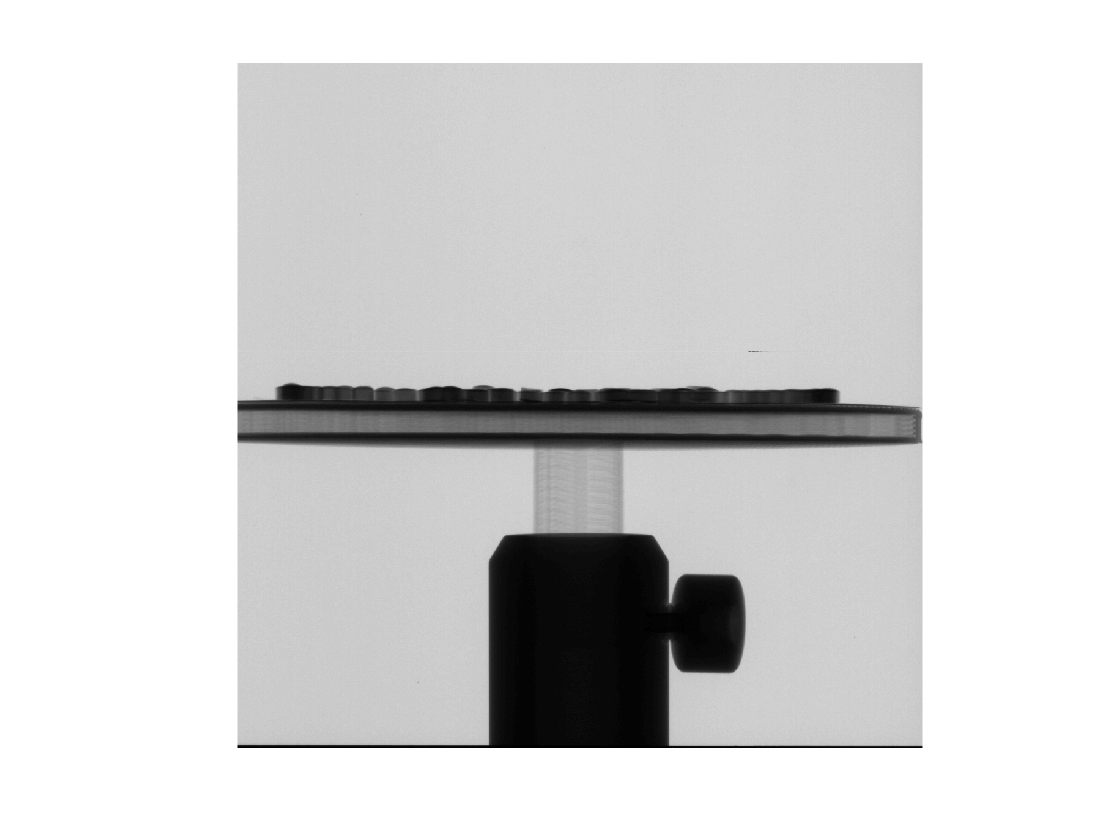}}
\put(240,98){\includegraphics[height=130pt]{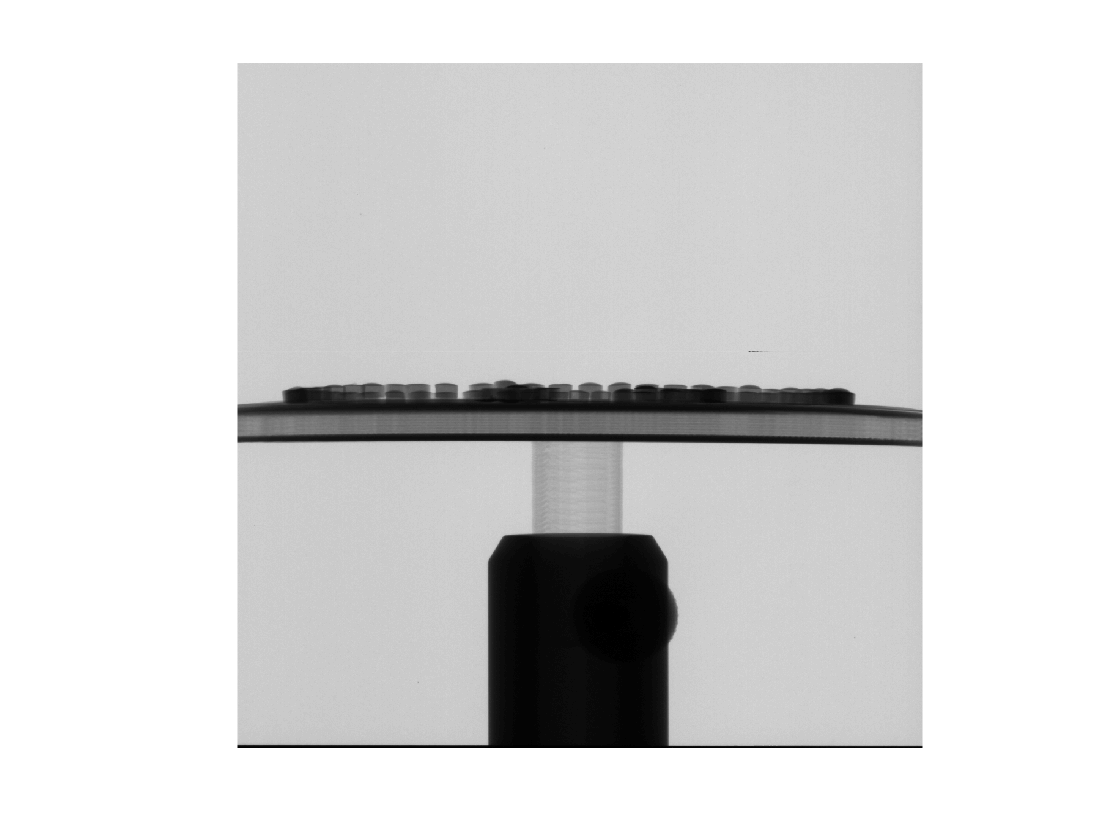}}
\end{picture}
\caption{\emph{Left}: Experimental setup used for collecting tomographic X-ray data. The target is attached to a computer-controlled rotator platform. \emph{Right}: Two examples of the resulting 2D projection images. The fan-beam data in the sinograms consist of the (down-sampled) central rows of the 2D projection images.}\label{fig:SetupAndProjections}
\end{figure}

The measurement setup was assembled in 2015 by Alexander Meaney as an MSc thesis project \cite{meaney2015design} and recently upgraded (1-3/2017). The setup is illustrated in Figure~\ref{fig:CTmachine} and the measurement geometry is shown in Figure~\ref{k1}. A set of 360 cone-beam projections with resolution $2240 \times 2368$ and the angular step of six (6) degree was measured. The exposure time was 1500 ms (i.e., one and a half second). The X-ray tube acceleration voltage was 50 kV and tube current 1 mA. See Figure~\ref{fig:SetupAndProjections} for two examples of the resulting projection images. 

\begin{figure}
\begin{picture}(390,390)
\put(195,415){\color{lightgray}\line(1,-3){128}}
\put(195,415){\color{lightgray}\line(-1,-3){128}}

\put(195,410){\color{yellow}\circle*{25}}
\put(145,80){\line(1,0){100}}
\put(145,105){\line(1,0){100}}
\put(145,130){\line(1,0){100}}
\put(145,155){\line(1,0){100}}
\put(145,180){\line(1,0){100}}
\put(145,80){\line(0,1){100}}
\put(170,80){\line(0,1){100}}
\put(195,80){\line(0,1){100}}
\put(220,80){\line(0,1){100}}
\put(245,80){\line(0,1){100}}
\put(152,165){$x_1$}
\put(152,140){$x_2$}
\put(155,112){$\vdots$}
\put(152,90){$x_N$}
\put(172,165){$x_{\scriptscriptstyle N+1}$}
\put(172,140){$x_{\scriptscriptstyle N+2}$}
\put(180,112){$\vdots$}
\put(174,90){$x_{2N}$}
\put(200,164){$\cdots$}
\put(200,139){$\cdots$}
\put(200,115){$\ddots$}
\put(200,89){$\cdots$}
\put(225,90){$x_{N^2}$}

\put(65,5){\color{red}\line(1,0){260}}
\put(65,30){\color{red}\line(1,0){260}}
\put(65,5){\color{red}\line(0,1){25}}
\put(90,5){\color{red}\line(0,1){25}}
\put(115,5){\color{red}\line(0,1){25}}
\put(140,5){\color{red}\line(0,1){25}}
\put(275,5){\color{red}\line(0,1){25}}
\put(300,5){\color{red}\line(0,1){25}}
\put(325,5){\color{red}\line(0,1){25}}
\put(70,13){$m_1$}
\put(95,13){$m_2$}
\put(120,14){$\cdots$}
\put(145,14){$\cdots$}
\put(255,14){$\cdots$}
\put(280,14){$\cdots$}
\put(305,13){$m_N$}
\end{picture}
\caption{The organization of the pixels in the sinograms {\tt m}\,=\,$[m_1,m_2,\ldots,m_{60K}]^T$ and reconstructions {\tt x}\,=\,$[x_1,x_2,\ldots,x_{N^2}]^T$with $N=128$. The picture shows the organization for the first projection; after that in the full angular view case, the target takes $6$ degree steps counter-clockwise (or equivalently the source and detector take $6$ degree steps clockwise) and the following columns of {\tt m} are determined in an analogous manner.}\label{fig:pixelDemo}
\end{figure}
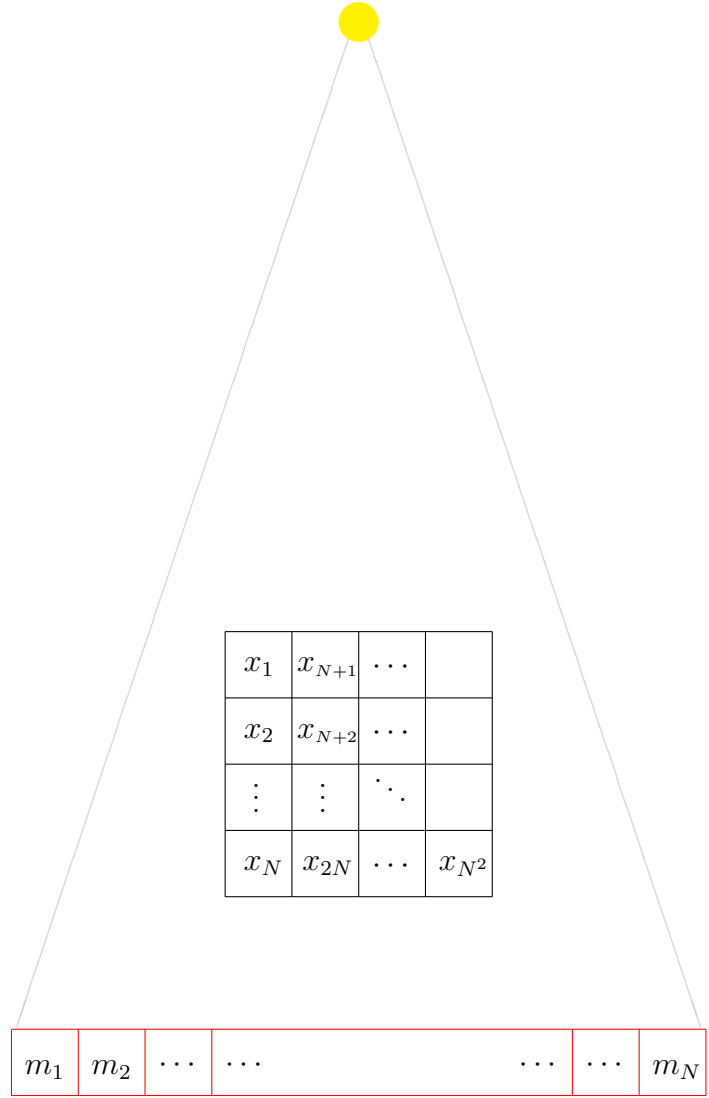

From the 2D projection images, the middle rows corresponding to the central horizontal cross-section of the emoji target were taken to form a
fan-beam sinogram of resolution $2240 \times 60$. These sinograms were further down-sampled by binning, taken logarithms and normalized to obtain the {\tt sinogram} in all the files specified in Section~\ref{sec:datasets}. The organization of the pixels in the sinograms and the reconstructions is illustrated in Figure~\ref{fig:pixelDemo}.

\begin{figure}
\begin{tikzpicture}[scale=2.0]
\draw [ultra thick,red] (2.5,-1.5) -- (2.5,1.5);
\draw[|-|, ultra thick] (-2.5,1.75) -- (1.5,1.75) node[below,midway]{FOD=540 mm}; 
\draw [|-|, ultra thick] (-2.5,-1.75) -- (2.5,-1.75)  node[below,midway]{FDD=630 mm}; 
\draw[|-|, ultra thick] (3.0,-1.5) -- (3.0,1.5); 
\draw[dotted] (-2.5,-1.75) -- (-2.5,1.75); 
\draw[dotted] (1.5,1.75) -- (1.5,0.0); 
\draw[dashed] (-2.5,0.0) -- (2.5,-1.5);
\draw[dashed] (-2.5,0.0) -- (2.5,1.5); 
\draw (3.65,0.0) node{W=120 mm}; 
\fill[thick] (1.5,0.0) circle (2pt);
\draw (1.5,0.25) node{COR};
\fill[thick, yellow] (-2.5,0.0) circle (3pt);
\end{tikzpicture}
\bigskip
\caption{Geometry of the measurement setup. Here FOD and FDD denote the focus-to-object distance and the focus-to-detector distance, respectively; the black dot COR is the center-of-rotation. The width of the detector (\ie{}, the red thick line) is denoted by W. The yellow dot is the X-ray source. To increase clarity, the $x$-axis and $y$-axis in this image are not in scale.}\label{k1}
\end{figure}
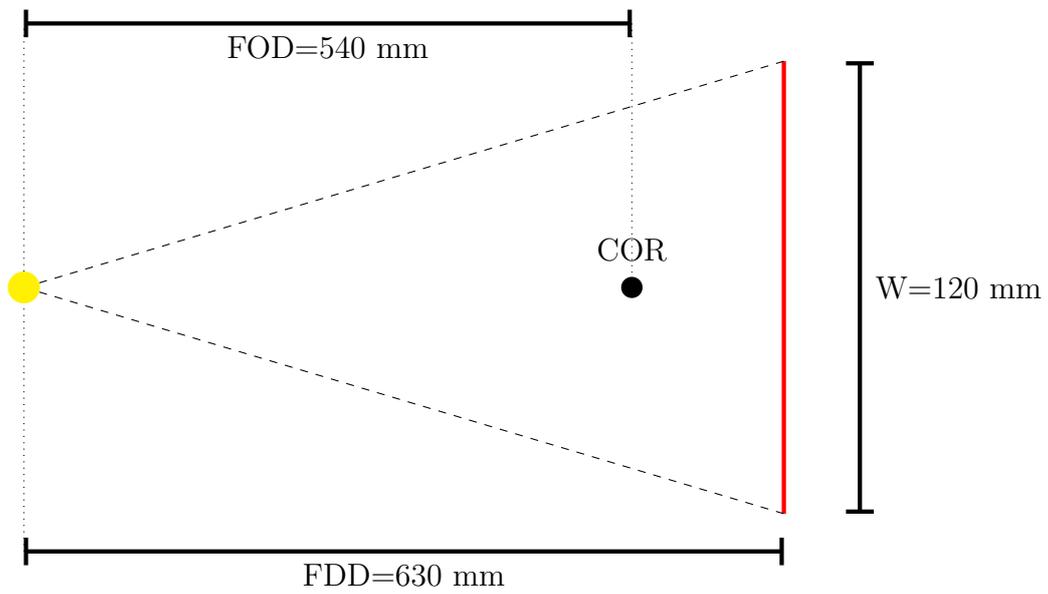

\section{Example of using the data}\label{sec:Demo}

The following MATLAB code demonstrates how to use the data. The code is also provided as the separate MATLAB script file {\tt example.m} and it assumes the data files (or in this case at least the file {\tt DataStatic\_128x60.mat}) are included in the same directory with the script file.

\begin{verbatim}
% Load the measurement matrix and the sinogram from
% file DataStatic_128x60.mat
load DataStatic_128x60 A sinogram

m = sinogram;
% Compute a Tikhonov regularized reconstruction using
% conjugate gradient algorithm pcg.m

% Define the number of slices
T = 33;
N     = sqrt(size(A,2)/T);
alpha = 10; % regularization parameter
fun   = @(x) A.'*(A*x)+alpha*x;
b     = A.'*m(:);
x     = pcg(fun,b);

% Compute a Tikhonov regularized reconstruction from only
% 10 projections
[mm,nn] = size(m);
ind     = [];
for iii=1:nn/6
    ind = [ind,(1:mm)+(6*iii-6)*mm];
end
m2    = m(:,1:6:end);
A     = A(ind,:);
alpha = 10; % regularization parameter
fun   = @(x) A.'*(A*x)+alpha*x;
b     = A.'*m2(:);
x2    = pcg(fun,b);

% Take a look at the sinograms of the first slice and the reconstructions
figure
subplot(2,2,1)
imagesc(m(:,1:size(m,2)/T))
colormap gray
axis square
axis off
title('Sinogram, 60 projections')
subplot(2,2,3)
imagesc(m2(:,1:size(m2,2)/T))
colormap gray
axis square
axis off
title('Sinogram, 10 projections')
% reshape the reconstruction
x = reshape(x,N,N,T);
x2 = reshape(x2,N,N,T);
% Show the first slice reconstruction
subplot(2,2,2)
imagesc(imrotate(x(:,:,1),-98,'bilinear','crop'))
colormap gray
axis square
axis off
title({'Tikhonov reconstruction,'; '60 projections'})
subplot(2,2,4)
imagesc(imrotate(x2(:,:,1),-98,'bilinear','crop'))
colormap gray
axis square
axis off
title({'Tikhonov reconstruction,'; '10 projections'})
\end{verbatim}

\clearpage
\section{3D reconstruction}
The video below is three-dimensional reconstruction of the target for {\tt DataStatic\_128x60.mat}) using the method implemented in Section~\ref{sec:Demo}. See \cite{barrett1994templates,Mueller2012}.

\bigskip
\bigskip
\includemedia[width=1\linewidth,height=1\linewidth,activate=pageopen,
passcontext,
transparent,
addresource=ReconsEmoji3DTime_size128_ang60.mp4,
flashvars={source=ReconsEmoji3DTime_size128_ang60.mp4}
]{\includegraphics[width=2\linewidth]{frame15c}}{VPlayer.swf}

\bigskip
For the reader's convenience, we also present some slices of the reconstruction below.

\begin{figure}
\begin{picture}(390,570)
\put(0,460){\includegraphics[width=140pt]{frame01c}}
\put(0,310){\includegraphics[width=140pt]{frame07c}}
\put(0,160){\includegraphics[width=140pt]{frame10c}}
\put(0,10){\includegraphics[width=140pt]{frame15c}}
\put(200,441){\includegraphics[width=225pt]{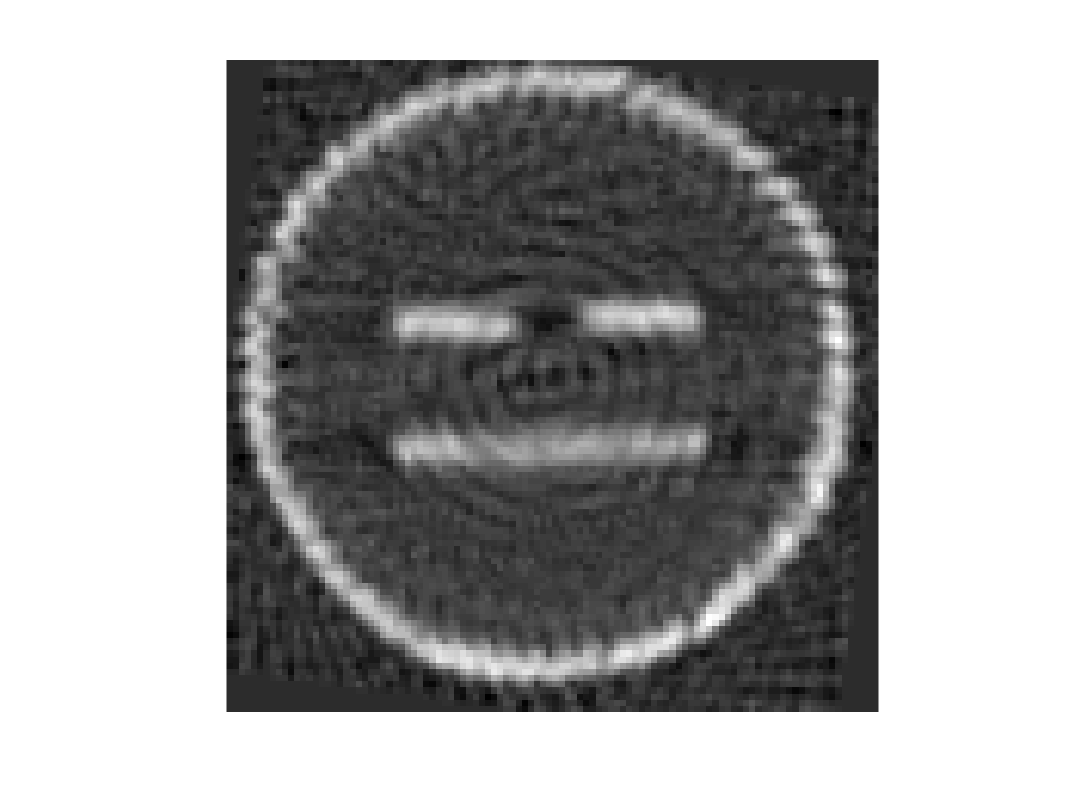}}
\put(200,295){\includegraphics[width=225pt]{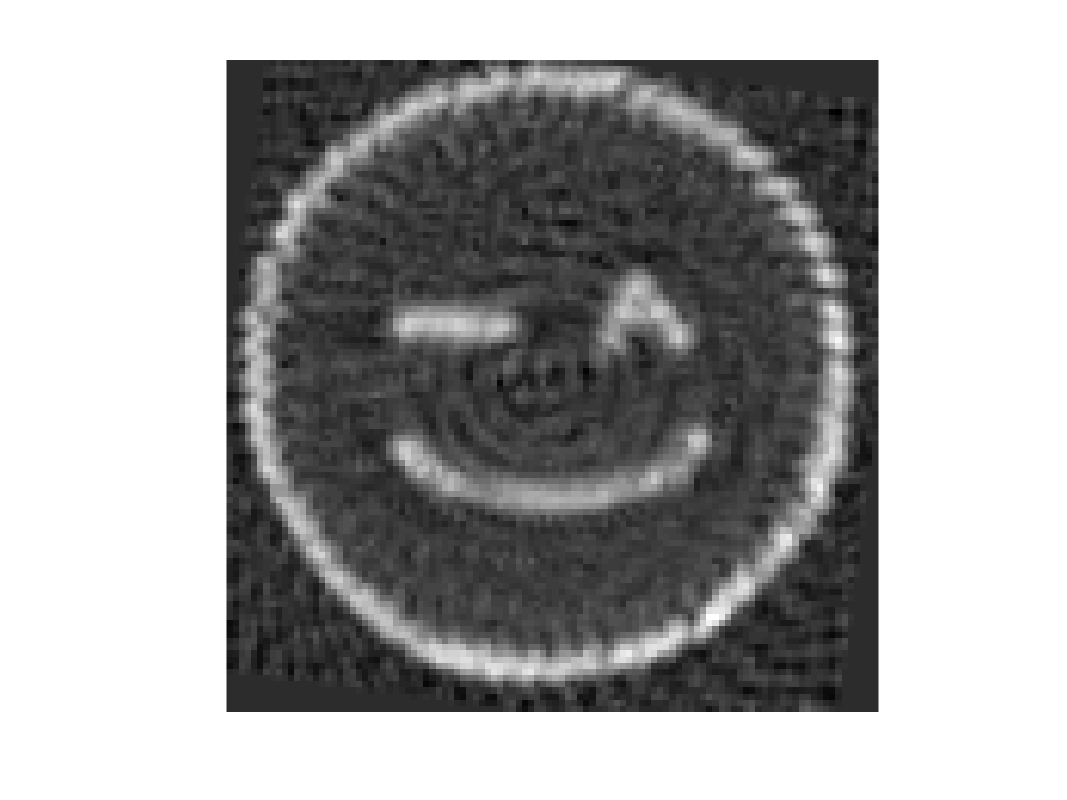}}
\put(200,143){\includegraphics[width=225pt]{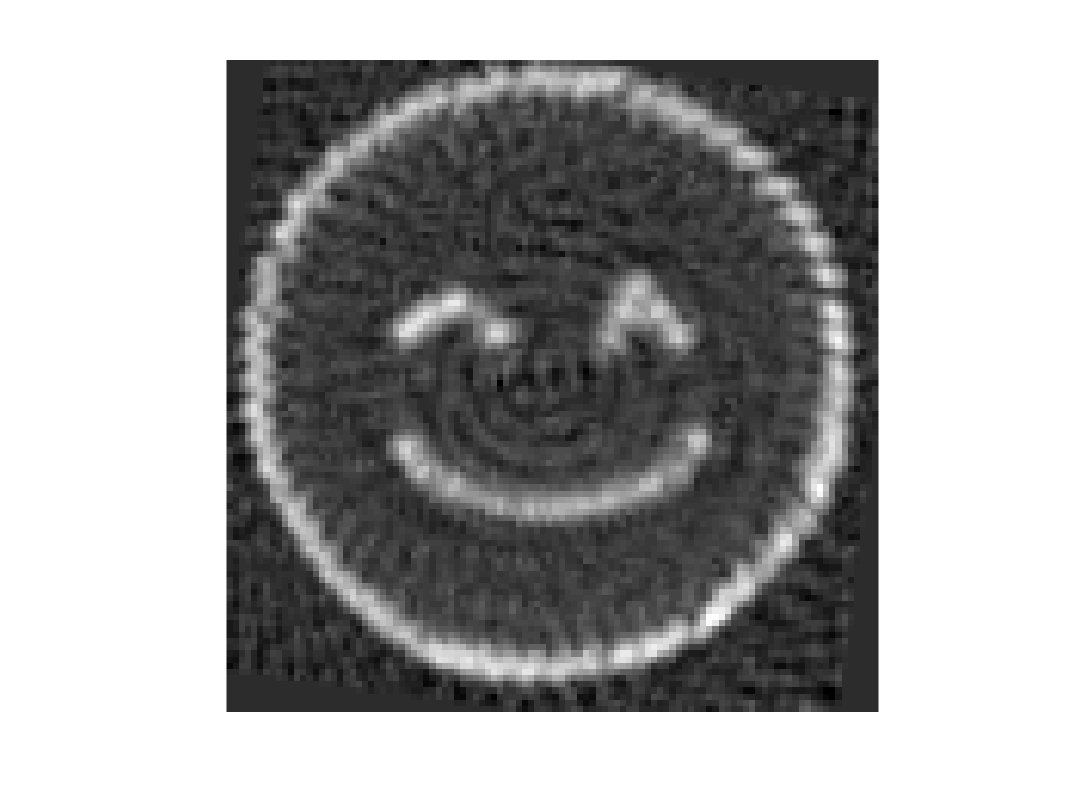}}
\put(200,-8){\includegraphics[width=225pt]{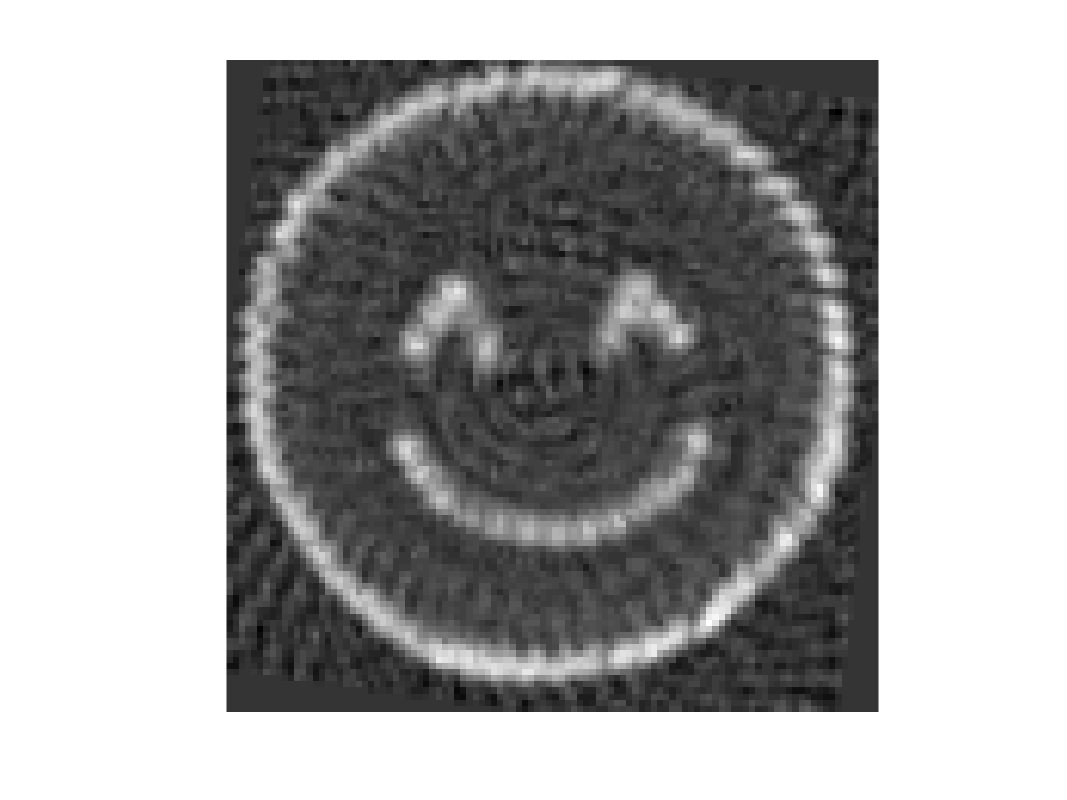}}
\caption{{\it Left}: Photographs of different 4 emoji faces. {\it Right}: Middle slices reconstructions of the corresponding photographs.}\label{fig:MiddleReconsEmoji}
\end{picture}
\end{figure}
\clearpage

\bibliographystyle{amsplain}
\bibliography{Inverse_problems_references}

\end{document}